\documentclass[fleqn,twoside]{article}
\usepackage{espcrc2}
\usepackage{latexsym}
\usepackage{graphicx}

\usepackage{graphicx}
\usepackage[figuresright]{rotating}

\newcommand{\AmS}{{\protect\the\textfont2
  A\kern-.1667em\lower.5ex\hbox{M}\kern-.125emS}}

\title{\vspace{-5.0cm} % this length needs adjustment
\begin{flushright}
{\normalsize Poster presented at ``Lattice 2002'' international
symposium, June 24-29, 2002, Cambridge, MA, USA}\\
\vspace{-0.2cm}
{\normalsize KEK-TH-846}\\
\vspace{-0.2cm}
{\normalsize RBRC-250}\\
\end{flushright}
\vspace*{2.5cm}
Nucleon axial charge from quenched lattice QCD with domain wall fermions and DBW2 gauge action}

\author{Shigemi Ohta\address[KEK]{Institute of Particle and Nuclear Studies, KEK,
Tsukuba, Ibaraki 305-0801, Japan}\address[RBRC]{RIKEN-BNL Research Center, Brookhaven National Laboratory, Upton, NY 11973, USA}\ %
for the RBC collaboration\thanks{Current members are Y.~Aoki, T.~Blum, N.~Christ, M.~Creutz, C.~Dawson, T.~Izubuchi, L.~Levkova, X.~Liao, G.~Liu, R.~Mawhinney, Y.~Nemoto, J.~Noaki, S.~Ohta, K.~Orginos, S.~Prelovsek, S.~Sasaki and A.~Soni.  We thank RIKEN, BNL and the U.S.\ DOE for providing the facilities essential for the completion of this work.}}
       
\begin{document}

\begin{abstract}
The domain wall fermion (DWF) method, with its almost perfectly
preserved chiral symmetry on the lattice, makes the calculation of the
nucleon axial charge particularly easy.  By maintaining chiral symmetry
and using the Ward-Takahashi (WT) identity, one has \(Z_A = Z_V\) and
the bare lattice calculation yields the physical value without explicit renormalization.  The
DBW2 improved gauge action provides further enhancement 
of the symmetry and hence a more accurate WT identity at coarse lattice spacing.  Taking advantage of these methods, we confirmed a significant volume
dependence of the nucleon axial charge on \((1.2 {\rm fm})^3\) and
\((2.4 {\rm fm})^3\) lattice volumes.
%The volume dependence of the tensor charge is also presented.
\vspace{1pc}
\end{abstract}

\maketitle

Domain wall fermions (DWF) \cite{DWF} preserve chiral symmetry on the
lattice by introducing a fictitious fifth dimension in which the symmetry
violation is exponentially suppressed.   The DBW2 (``doubly blocked Wilson
2'') improved gauge action \cite{QCDTARO} improves the approach to the
continuum by adding rectangular (2\(\times\)1) Wilson loops to the action.  By
combining the two, the ``residual mass,'' which controls low energy chiral
behavior, is driven to as small as \(am_{\rm res} < O(10^{-4}) \), or \(\ll
{\rm 1\,MeV}\).  Successful numerical lattice calculations investigating chiral
symmetry and the ground-state hadron mass spectrum \cite{DWFspectroscopy}, Kaon matrix elements \cite{DWFKaon}, and negative-parity excited nucleon (\(N^*\)) mass \cite{DWFNstar} have been reported.  Here we apply the method to the axial charge, \(g_{_A}\), the simplest of nucleon electroweak matrix elements.

From neutron \(\beta\) decay, we know \(g_{_V} = G_F \cos \theta_c\) and \(g_{_A}/g_{_V} = 1.2670(35)\) \cite{PDGgA}.  These quantities are defined as \(q^2\rightarrow 0\) limits of the form factors \(g_{_V}(q^2)\) and \(g_{_A}(q^2)\) in the relevant vector and axial current matrix elements, \(\displaystyle \langle n| V^-_\mu(x) | p \rangle\),
\[
i\bar{u}_n [\gamma_\mu g_{_V}(q^2)
             +q_\lambda \sigma_{\lambda\mu} g_{_M}(q^2) ] u_p e^{-iqx},
\]
and \(\displaystyle \langle n| A^-_\mu(x) | p \rangle\),
\[
i\bar{u}_n \gamma_5
             [\gamma_\mu g_{_A}(q^2)
             +q_\mu g_{_P}(q^2) ] u_p e^{-iqx}.
\]
On the lattice, in general, we calculate the relevant matrix elements of
these currents with a lattice cutoff, \(a^{-1}\) \(\sim\) 1-2 GeV, and extrapolate to the continuum, \( a \rightarrow 0\), introducing lattice renormalization: \(g_{_V, _A}^{\rm ren} = Z_{_V,
_A}^{\rm lat} g_{_V, _A}^{\rm lat}\).  Also, unwanted lattice artefacts may result in unphysical mixing of chirally distinct operators.  DWF make the calculation particularly easy, because the chiral symmetry is almost exact; the relation \(Z_{_A} = Z_{_V}\) is maintained, so the lattice ratio \((g_{_{A}}/g_{_{V}})^{\rm lat}\) directly yields the renormalized ratio.

It is interesting to note the pre-QCD model estimations for this quantity.
The non-relativistic quark model in its simplest form gives \(5/3\), and the
MIT bag model gives 1.07.  The lattice QCD calculations with Wilson or clover
fermions underestimate $g_A$ by 10 to 25 \% \cite{WilsongA}.  These calculation also suffer from systematic errors arising from  \(Z_{_A} \neq Z_{_V}\) and other renormalization complications.

Our formulation follows the standard one, and was summarized in our Lattice 2001 proceedings \cite{Lat01}.  In addition to the wall sources, sequential sources are added  \cite{Kostas}.

Numerical calculations with Wilson (single plaquette) gauge action were
performed at \(\beta=6.0\) on a \(16^3\times 32 \times 16\) lattice and \(M_5
= 1.8\) with nucleon source at \(t=5\), sink at 21, and current insertions in
between.  Results from 400 gauge configurations are summarized as follows.
The local vector current renormalization \(Z_{_V} = 1/g_{_V}^{\rm lattice}\)
is well-behaved: the value \(0.764(2)\) at \(m_f = 0.02\) agrees well with
\(Z_{_A} = 0.7555(3)\) of the corresponding local axial current
renormalization \cite{DWFspectroscopy,DWFNPR}. A linear fit gives \(Z_{_V} =
0.760(7)\) at \(m_f = 0\), and a quadratic fit, 0.761(5).  The axial to vector charge ratio, \(g_{_A}/g_{_V}\), averaged in \(10\le t \le 16\) time slices, yields a value of 0.81(11) in the linearly extrapolated chiral limit \(m_f=0\).  These are essentially the same as last year \cite{Lat01}, but have been confirmed with much higher accuracy obtained through higher statistics.

Possible reasons for this small value of  \(g_{_A}/g_{_V}\) are finite lattice volume \cite{JaffeCohen}, excited states (small separation between \(t_{\rm source}\) and \(t_{\rm sink}\)), and quenching (zero modes, absent pion cloud, and so on).  Here we concentrate on the first possibility.  To investigate size-dependence, we need good chiral behavior or sufficient proximity  to the continuum, and big enough lattice spatial volume at the same time.  Improved gauge actions help both \cite{DWFDBW2}.  In particular the DBW2 action \cite{QCDTARO}  gives
very small residual chiral symmetry breaking, \(am_{\rm res} < 10^{-3}\), at a
lattice spacing of  \(a\) \(\sim\) 0.15 fm.  At this lattice cutoff the hadron mass spectrum is well-behaved in the chiral limit too: \(am_\rho = 0.592(9)\), \(m_\rho/m_N \sim 0.8\), and \(m_\pi (m_f=0.02) \sim 0.3 a^{-1}\).

DBW2 calculations are performed with both wall and sequential sources on two lattices:
\(8^3 \times 24 \times 16\) (\(\sim\) \((1.2 {\rm fm})^3\)), 405 configurations (wall and sequential), and \(16^3 \times 32 \times 16\) (\(\sim\) \((2.4 {\rm fm})^3\)), 100 configurations  (wall) and 204 (sequential).  Nucleon  source-sink separation is set at about 1.5 fm, and quark mass values of
\(m_f\) = 0.02, 0.04, 0.06, 0.08 and 0.10 are used.  So the pion mass is as
low as \(m_\pi \ge 390 {\rm MeV}\) and satisfies \(m_\pi L\ge 4.8\) and 2.4, respectively.  The residual mass at this cutoff is about 0.8 MeV or \(a m_{\rm res} \sim 6\times 10^{-4}\).

\begin{figure}[t]
\vspace{9pt}
\includegraphics[width=\columnwidth]{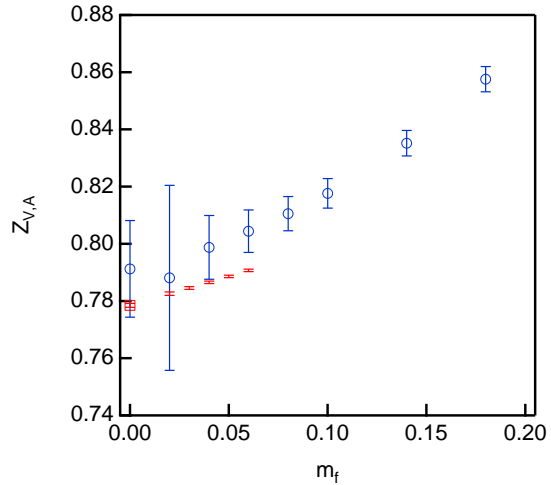}
\caption{Local current renormalization factors, vector (\(Z_{_V}\), circle) and axial (\(Z_{_A}\), square).}
\label{fig:DBW2Z}
\end{figure}

Again the vector and axial current renormalization factors are well-behaved (see Figure \ref{fig:DBW2Z}.)
The vector renormalization, \(Z_{_V}\), calculated as \(1/g_{_V}\), shows slight quadratic dependence on \(m_f\) as expected: \(V_\mu^{\rm conserved} = Z_{_V} V_\mu^{\rm local} + {\cal O}
(m_f^2 a^2)\), yielding a value \(Z_{_V} = 0.791(17)\) in the chiral limit.  This
agrees well with the axial renormalization, \(Z_{_A} = 0.77759(45)\), obtained from
\(\langle A_{_\mu}^{\rm conserved} (t) [\bar{q}\gamma_5 q](0) \rangle =
Z_{_A} \langle A_\mu^{\rm local} (t) [\bar{q}\gamma_5 q](0) \rangle\) \cite{DWFNPR}.  This assures that the lattice calculation of the axial-to-vector ratio directly yields its continuum value without any further renormalization: \((g_{_A}/g_{_V})^{\rm lat} = (g_{_A}/g_{_V})^{\rm ren}\).

\begin{figure}[t]
\vspace{9pt}
\includegraphics[width=\columnwidth]{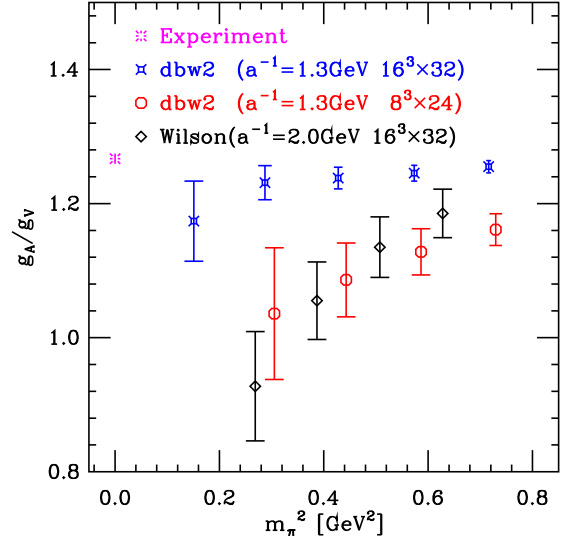}
\caption{\((g_{_A}/g_{_V})^{\rm ren}\): \(m_f\) and volume dependence in physical scale set by \(\rho\) meson mass.  Wilson (diamond) and DBW2 results are consistent in the smaller volumes.}
\label{fig:ga_o_gv_phys}
\end{figure}

In Figure \ref{fig:ga_o_gv} the quark-mass dependence of the ratio obtained from sequential source calculations is summarized for the two lattice volumes:
\begin{figure}[t]
\vspace{9pt}
\includegraphics[width=\columnwidth]{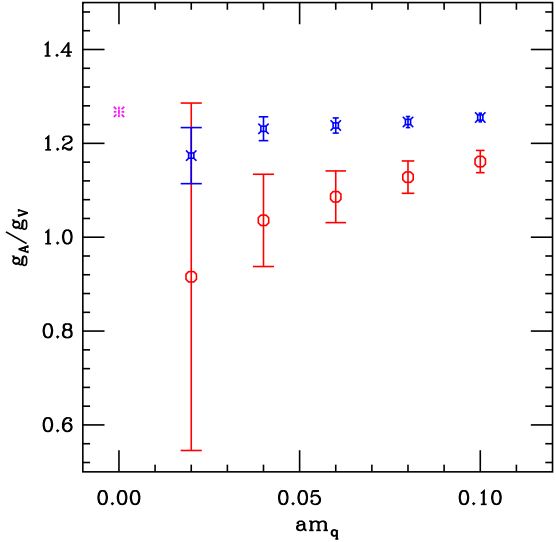}
\caption{\((g_{_A}/g_{_V})^{\rm ren}\): \(m_f\) and volume dependence.  Sequential source results on a large (cross) and small (circle) lattice.  Wall source results are consistent with larger errors.}
\label{fig:ga_o_gv}
\end{figure}
Clear volume dependence is seen between \((2.4 {\rm fm})^3\) and \((1.2 {\rm fm})^3\) volumes.
The large volume results with the sequential source show a very mild \(m_f\) dependence and
extrapolate to a value of \((g_{_A}/g_{_V})^{\rm ren} = 1.21(3)(+4)\).  The first error is statistical and the second is systematic from a preliminary finite-size analysis.
At the lightest quark mass value of \(m_f\)=0.02, the volume dependence is no
longer statistically significant.  Indeed at this \(m_f\) even the larger
volume may not be big enough.  In Figure \ref{fig:ga_o_gv_phys} the same data
are presented in physical units together with the small volume Wilson gauge action results.
The smaller volume results with Wilson (\(16^3\times 32\) at \(a\) \(\sim\) 0.1 fm) and DBW2 (\(8^3\times 24\) at \(a\) \(\sim\) 0.15 fm) actions are mutually consistent.  In summary we confirmed a lattice-volume dependence of the nucleon axial charge, \(g_{_A}/g_{_V}\).
Similar dependence was seen in the bare tensor charge.
Further investigation is under way.


\begin{thebibliography}{99}
\bibitem{DWF} D.B.\ Kaplan, Phys.\ Lett.\ B288, 342 (1992); Y.\ Shamir, Nucl.\ Phys.\ B406, 90 (1993); V.\ Furman and Y.\ Shamir, Nucl.\ Phys.\ B439, 54 (1995).
\bibitem{QCDTARO} P.\ de Forcrand {\it et al.}\ (QCD-TARO collaboration), Nucl.\ Phys.\ B577, 263 (2000).
\bibitem{DWFspectroscopy} T.\ Blum, {\it et al.}\ (RBC collaboration), to appear in Phys.\ Rev.\ D, hep-lat/0007038.
\bibitem{DWFKaon} T.\ Blum, {\it et al.}\ (RBC collaboration), RBRC Scientific Articles 4; hep-lat/0110075, submitted for publication in Phys.\ Rev.\ D.
\bibitem{DWFNstar} S.\ Sasaki, T.\ Blum and S.\ Ohta  (for RBC collaboration), Phys.\ Rev.\ D65, 074503 (2002).
\bibitem{PDGgA} The Particle Data Group.
\bibitem{WilsongA} M.\ Fukugita, Y.\ Kuramashi,
M.\ Okawa and A.\ Ukawa, Phys.\ Rev.\ Lett.\ 75, 2092 (1995); K.F.\ Liu, S.J.\ Dong, T.\
Draper and J.M.\ Wu, Phys.\ Rev.\ D49, 4755 (1994); M.\ G\"ockeler {\it et al.}, Phys.\
Rev.\ D53, 2317 (1996); D.\ Dolgov {\it et al.}, hep-lat/0201021; S.\ Capitani {\it et al.}, Nucl.\ Phys.\ B (Proc.\ Suppl.)\ 79, 548 (1999); R.\ Horsley {\it et al.}, Nucl.\ Phys.\ B (Proc.\ Suppl.)\ 94, 307 (2001); S.\ G\"usken {\it et al.}, Phys.\ Rev.\ D59, 114502 (1999).
\bibitem{Lat01} S. Sasaki, Nucl.\ Phys.\ Proc.\ Suppl.\ 106, 30 (2002); and references cited there in.
\bibitem{Kostas} K.\ Orginos for the RBC collaboration, in these proceedings.
\bibitem{DWFNPR} T.\ Blum {\it et al.}\ (RBC Collaboration), Phys.\ Rev.\ D66, 014504 (2002).
\bibitem{JaffeCohen} R.L.\ Jaffe, Phys.\ Lett.\ B529, 105 (2002); T.D.\ Cohen, Phys.\ Lett.\ B529, 50 (2002).
\bibitem{DWFDBW2} RBC collaboration, in preparation.
\end{thebibliography}
\end{document}